\documentclass[aps,pre,twocolumn,unsortedaddress,showpacs,floatfix,10pt,letterpaper]{revtex4-1}

\usepackage{graphicx}    % Include figure files
\usepackage{dcolumn}     % Align table columns on decimal point
\usepackage{bm}          % bold math
\usepackage{color}       % Color text
\usepackage{multirow}
\usepackage{enumitem}
\usepackage{units}
\usepackage{url}
\usepackage{amsmath}

\begin{document}
\title{A histogram-free multicanonical Monte Carlo algorithm for the
  basis expansion of density of states}
%\subtitle{}
%\subtitlenote{}

\author{Ying Wai Li}
\email[]{yingwaili@ornl.gov \\
This manuscript has been authored by UT-Battelle, LLC under
  Contract No. DE-AC05-00OR22725 with the U.S. Department of
  Energy. The United States Government retains and the publisher, by
  accepting the article for publication, acknowledges that the United
  States Government retains a non-exclusive, paid-up, irrevocable,
  worldwide license to publish or reproduce the published form of this
  manuscript, or allow others to do so, for United States Government
  purposes. The Department of Energy will provide public access to
  these results of federally sponsored research in accordance with the
  DOE Public Access Plan
  (http://energy.gov/downloads/doe-public-access-plan).}
\affiliation{National Center for Computational Sciences, Oak Ridge
  National Laboratory, Oak Ridge, Tennessee 37831, U.S.A.}

\author{Markus Eisenbach}
\affiliation{National Center for Computational Sciences, Oak Ridge National Laboratory, Oak Ridge, Tennessee 37831, U.S.A.}

\begin{abstract}
We report a new multicanonical Monte Carlo (MC) algorithm to obtain the 
density of states (DOS) for physical systems with continuous state variables 
in statistical mechanics.  Our algorithm is able to obtain an analytical form 
for the DOS expressed in a chosen basis set, instead of a numerical array 
of finite resolution as in previous variants of this class of MC methods such as
the multicanonical (MUCA) sampling and Wang-Landau (WL) sampling. This 
is enabled by storing the visited states directly in a data set and 
avoiding the explicit collection of a histogram. This practice also
has the advantage of avoiding undesirable artificial errors caused by
the discretization and binning of  continuous state variables. Our
results show that this scheme is capable of obtaining converged
results with a much reduced number of Monte Carlo steps, leading to a
significant speedup over existing algorithms.
\end{abstract}


\maketitle

\section{Introduction}

Monte Carlo (MC) methods are one of the major computational techniques
in statistical physics for the study of finite temperature properties and
thermodynamics of materials \cite{Landau_Binder}. Traditional MC
methods such as the Metropolis algorithm \cite{Metropolis1953}, an
importance sampling method, works by generating a Markov chain of
energy states $E$ that obey the Boltzmann distribution, $e^{-E/k_B
  T}$, which describes the probability of finding the system at a certain
energy state at a given temperature $T$. Thermodynamics properties are
then calculated by averaging over the entire Markov chain after
equilibration. A well-known limitation of the Metropolis method is the
``critical slowing down'' near phase transitions \cite{Hohenberg1977},
where the correlation time diverges at the critical temperature
$T_C$. Hence, simulations around and below $T_C$ are simply impractical or
unreliable to perform.

Important breakthroughs were introduced by advanced techniques
such as the reweighting methods, which allow for the procurement of a
distribution function of properties. They can be used to obtain
properties at a temperature other than the simulation temperature by
``reweighting'' the distribution function properly. Umbrella sampling
\cite{BENNETT1976,Torrie1977}, multihistogram method
\cite{Ferrenber1989}, multicanonical (MUCA) sampling
\cite{berg_neuhaus_1991,berg_neuhaus_1992}, and more recently
Wang-Landau sampling \cite{Wang2001b,Wang2001a}, all belong
to this class of reweighting methods. Because of a special formulation
of the sampling weights that control the acceptance probability, the
random walks in these methods are not ``trapped'' in local minima as
in Metropolis sampling. They are thus able to circumvent the critical
slowing down problem. Among the reweighting methods, Wang-Landau
sampling is proven to be quite robust because the simulation is
performed \textit{independent} of temperature. The resulting distribution
function is essentially the density of states (DOS) or
the energy degeneracy of the system. Thus it reflects only the intrinsic
properties defined by the Hamiltonian. The DOS allows for the direct
access to the microcanonical entropy, with which all the
thermodynamics properties including the specific heat and free energy
can be calculated. This feature is essential to enable a reliable
study of phase transitions and critical phenomena, particularly at low
temperatures. 

With the advancement of high performance computers (HPC), it is now possible
to combine Wang-Landau sampling with first-principles methods,
e.g. density functional theory (DFT) \cite{hohenberg_1964,kohn_1965},
to simulate finite temperature materials properties to a high accuracy
that is comparable with experimental observations
\cite{Eisenbach2009a,Khan2016}. However, first-principles energy
calculations are computationally intensive; and yet a reliable
Wang-Landau sampling often needs a minimum of millions of MC steps
(i.e. energy calculations) for one single simulation. The time required to
finish a simulation is often measured in weeks or even months on one of the
fastest supercomputers currently available. Such a huge
computational cost is barely affordable. The type of scientific
problems that can be practically solved by this approach are, for this
reason, still very limited. 

To address this problem, improvements of existing Monte Carlo
algorithms are required. In general, two feasible strategies are available: one
is the parallelization of existing algorithms, in which computational
cost is spread over multiple computing units. Examples include
parallel tempering \cite{Swendsen1986,Hukushima1996}, parallel
Wang-Landau sampling on a graphical processing unit (GPU)
\cite{Yin2012}, replica-exchange Wang-Landau sampling
\cite{Vogel2013,Vogel2014}, and parallel multicanonical sampling
\cite{Zierenberg2013}. Another strategy is to find ways to reduce the
number of MC steps needed to complete a simulation. This is normally
done by introducing tricks within the framework of existing
algorithms; but the number of MC steps saved is often small.

In this paper, we present a new multicanonical Monte Carlo algorithm
that takes both strategies into account. Our scheme is readily
parallelizable to exploit the power of current HPC architectures. In
addition, our algorithm is able to attain comparable accuracy with
Wang-Landau sampling, using only about 1/10 of the number of MC
steps. This order of magnitude of reduction in the number of energy
evaluations is particularly crucial when first-principles methods are
employed for calculating the energy. Moreover, for the very first
time, our algorithm provides a viable means to obtain the density of
states in an analytical form. This algorithm will be particularly
useful to fit the functional form of the density of states to aid
theoretical studies.

\section{Description of the algorithm}

\subsection{An overview}
Our novel algorithm is inspired by previous multicanonical (MUCA)
\cite{berg_neuhaus_1991,berg_neuhaus_1992} and Wang-Landau
(WL)\cite{Wang2001a,Wang2001b} Monte Carlo methods. Therefore our
algorithm shares many of its underlying principles with these earlier
methods. The major advantage of our scheme over the previous ones is
that our algorithm, for the first time, provides a viable avenue to
estimate an \textit{analytic form} of the density of states in
energy, denoted by $g(E)$. Here $E$ stands for an energy the simulated
physical system can realize. We assume an analytic form for the
natural log of $g(E)$ in terms of an orthonormal basis set
$\{\phi_i(E)\}$ each weighted by the coefficient $g_i$:
\begin{equation}
\label{g(E)}
\ln g(E) = \sum_{i=1}^{N}g_i\phi_i(E),
\end{equation}
with $N$ being the number of basis functions utilized in the
expansion. The estimation of $g(E)$ will be improved iteratively
later during the course of the simulation by a similarly defined, yet
slightly modified, correction function $c(E)$:
\begin{equation}
\label{correction1}
\ln c(E) = \sum_{i=1}^{N}c_i\phi_i(E),
\end{equation}
where $c_i$ is the weighting coefficient for $\phi_i(E)$ in the
correction.

The algorithm begins with an initial guess of $\tilde{g}(E) = 1$
(i.e., $\ln \tilde{g}(E) = 0$). In other words, it is a uniform
distribution with no energy degeneracy. Next, a series of
Monte Carlo moves is performed and a Markov chain of $k$ energies is 
generated to construct a data set $\mathcal{D}=\{ E_1, E_2, ..., E_j,
..., E_k \} $ according to the following acceptance probability:
\begin{equation}
p(E_j \to E_{j+1}) = \min\left( \frac{\tilde{g}(E_j)}{\tilde{g}(E_{j+1})} , 1 \right ).
\end{equation}
Note that the acceptance rule follows that of the Wang-Landau
algorithm\cite{Wang2001a}. That is, if the trial energy $E_{j+1}$ is
rejected, the previous accepted state of the system should be recovered,
but the associated energy $E_j$ would be counted again as $E_{j+1}$. A
Monte Carlo move is then performed on the reverted state to generate
the next trial energy $E_{j+2}$.

After the data set $\mathcal{D}$ is generated, it is used to find the
correction $c(E)$ that improves the estimated density of states
$\tilde{g}(E)$ such that:
\begin{equation}
\label{updateDOS}
\ln \tilde{g}(E) \to \ln \tilde{g}(E) + \ln c(E).
\end{equation}
The details of obtaining the correction function $c(E)$ from the data set
$\mathcal{D}$ will be further described below in subsection
\ref{correction}. For now assume that we have updated the estimated
density of states $\tilde{g}(E)$ with $c(E)$ using
Eq. (\ref{updateDOS}). The simulation is then brought to the next
iteration with $\mathcal{D}$ and $\ln c(E)$ reset to empty or zero,
respectively, while $\tilde{g}(E)$ will be kept unchanged and carried
over to the next iteration as the new sampling weights. The process of
generating the data set $\mathcal{D}$ and obtaining the correction
$c(E)$ is then repeated. The iteration repeats and terminates when
$\ln c(E) \rightarrow 0$. In this case the DOS becomes a fixed point
of the iterative process and convergence is reached.

\subsection{Obtaining the correction $c(E)$ from data set
  $\mathcal{D}$}
\label{correction}

The key of the above framework is to obtain an analytic
expression for the correction $c(E)$, or $\ln c(E)$ in the actual
implementation of our algorithm. To do
so, we must first obtain an analytic expression for the empirical
cumulative distribution function (ECDF) of the data $\mathcal{D}$,
from which $c(E)$ can be deduced. 

\subsubsection{Obtaining an analytic expression for the empirical
cumulative distribution function (ECDF)}

We construct the ECDF following the scheme proposed by Berg
and Harris \cite{berg_data_2008}, which we outline here. Recall that
our data set $\mathcal{D}$ is a collection of $k$ energies generated
from a Monte Carlo Markov chain. The energies are first rearranged in
ascending order: 
\begin{equation}
\begin{aligned}
\mathcal{D} &= \{ E_1, E_2, ..., E_j, ..., E_k \} \\
                    &= \{ E_{\pi_1}, E_{\pi_2}, ..., E_{\pi_j}, ..., E_{\pi_k} \},
\end{aligned}
\end{equation}
where $\pi_1$, ..., $\pi_k$ is a permutation of 1, ..., $k$ such that
$E_{\pi_1} \leq E_{\pi_2} \leq ... \leq E_{\pi_j} \leq ... \leq E_{\pi_k}$. The empirical
cumulative distribution function (ECDF) is then defined as:
\begin{equation}
\bar{F}(E) = \frac{j}{k}  \hspace{5mm}\textrm{for $E_{\pi_j} \le E < E_{\pi_k}$ }.
\end{equation}

\noindent
 Assuming that the ECDF can be decomposed into two components:
\begin{equation}
\label{ECDF}
\bar{F}(E) = F_0(E) + \bar{R}(E),
\end{equation}
where $F_0(E) = (E-E_{\pi_1}) / (E_{\pi_k}-E_{\pi_1})$ is a straight line for
$E \in [E_{\pi_1}, E_{\pi_k} ]$, and $\bar{R}(E)$ defines the
empirical remainder. The choice of $F_0(E)$ as a straight line is
based upon the following observations: for traditional histogram
methods, the ECDF plays the role of the cumulative histogram that can
be constructed directly from the histogram $H(E)$. Nevertheless, the
ECDF does not suffer from the
binning effect. The derivative of ECDF is then equivalent to the
histogram in traditional methods: $H(E) = d\bar{F}(E) / dE$. In such
schemes, obtaining a ``flat'' histogram is an indicator that the
energy space is being sampled uniformly. The sampling weights are
continuously adjusted to direct the random walk from highly accessible
states to rare events, either periodically in MUCA or adaptively in
WL, to achieve this goal. Here, a ``flat histogram'' is equivalent to
an ECDF with a straight line of a constant slope.

The next task would be finding an analytic expression for the
remainder $R(E)$ to fit the empirical data $\bar{R}(E)$. $R(E)$
signifies the deviation from the ideal (uniform) sampling, which
will inform us on how to amend the weights to drive the random
walks. It is expected that $R(E)$ will be related to the correction
$c(E)$. Therefore, it is reasonable to assume that $R(E)$ can be
similarly expanded in terms of an orthonormal basis set $\{\psi_i(E)\}$:
\begin{equation}
\label{R(E)}
R(E) = \sum_{i=1}^{m}r_i\psi_i(E),
\end{equation}
where $m$ is the number of terms in the expression. The coefficients
$r_i$ can be then be found by:
\begin{equation}
\label{remainder_coef}
\begin{aligned}
r_i &= \mathcal{N} \int_{E_{\pi_1}}^{E_{\pi_k}} R(E) \psi_i(E) dE,
\end{aligned}
\end{equation}
with $\mathcal{N}$ being a normalization constant dependent on the
choice of the basis set $\{\psi_i(E)\}$. Note also that the basis set
$\{\psi_i(E)\}$ needs to be able to satisfy the ``boundary condition''
at $E_{\pi_1}$ and $E_{\pi_k}$ that $R(E_{\pi_1}) = R(E_{\pi_k}) = 0$
by definition. Since $R(E)$ is indeed an empirical function resulted
from the ECDF, the integral in Eq. (\ref{remainder_coef}) is a quick
summation for the area under curve. 

The remaining question is to determine the number of terms $m$ in
Eq. (\ref{R(E)}) to fit $\bar{R}(E)$. This is done by an iterative
procedure starting from $m = 1$ where there is only one term in the
sum. A statistical test is then performed to measure the probability
$p$ that this $R(E)$ is a ``good'' fit to $\bar{R}(E)$. That is, $p$
is the probability of obtaining the empirical remainder $\bar{R}(E)$
if the data is generated according to the distribution specified by
$R(E)$. We follow the suggestion of \cite{berg_data_2008} and use the
Kolmogorov-Smirnov test, but other statistical tests for arbitrary
probability distributions can also be used. If $p < 0.5$, we increase
$m$ to $m+1$ and repeat the statistical test, until $p \geq 0.5$ is
reached. The number of terms $m$ is then fixed at this point.  Note
that in principle, increasing $m$ further would result in a ``better
fit'' and thus a larger $p$. However, it is not preferable because it
increases the risk of over-fitting a particular data set and would be
difficult to correct through latter iterations. Thus we choose the
criterion $p \geq 0.5$ to keep the expression as simple as possible,
and to maintain some levels of stability against noise.

With the expression of $R(E)$, the analytic approximation of ECDF
can then be obtained:
\begin{equation}
\label{F(E)}
F(E) = F_0(E) + R(E).
\end{equation} 

\subsubsection{From ECDF $F(E)$ to the correction $c(E)$}
                                                    
Finally, the expression of $F(E)$ in Eq. (\ref{F(E)}) is used to
obtain the correction $c(E)$ (or $\ln c(E)$ in practice). Recall the
definition of the cumulative distribution function (CDF) for a
continuous variable, which can be constructed from the probability
density function. They are, respectively, equivalent to $F(E)$ and
$H(E)$:
\begin{equation}
\label{CDF}
F(E) = \int_{-\infty}^{E} H(E') dE'.
\end{equation} 

Combining Eqs. (\ref{F(E)}) and (\ref{CDF}) and taking derivatives
from both sides to obtain $H(E)$ yields:
\begin{equation}
\begin{aligned}
\label{H(E)}
H(E) = \frac{dF(E)}{dE} &= \frac{dF_0(E)}{dE} + \frac{dR(E)}{dE} \\
                                   &= \frac{1}{E_{\pi_k}-E_{\pi_1}} + \sum_{i=0}^{m}r_i \frac{d\psi_i(E)}{dE}.
\end{aligned}
\end{equation}

As in traditional multicanonical sampling methods, the histogram
$H(E)$ is used to update the estimated density of states $\tilde{g}(E)$,
hence the sampling weights for the next iteration. Observe that the
first term in Eq. (\ref{H(E)}) is just a constant independent of the
value of $E$, one can safely omit it in the correction. Thus,
\begin{equation}
\begin{aligned}
\label{correction2}
\ln c(E) &= \sum_{i=1}^{m}r_i \frac{d\psi_i(E)}{dE},
\end{aligned}
\end{equation}
which has the same form as Eq. (\ref{correction1}) with 
\begin{equation*}
c_i \phi_i(E) = r_i \frac{d\psi_i(E)}{dE} \textrm{ and } N = m. 
\end{equation*}
Finally, the estimated density of states $\tilde{g}(E)$ is updated
using Eq. (\ref{updateDOS}).

\section{Test case: numerical integration}

The algorithm was originally designed with the motivation of sampling
physical systems with a continuous energy domain. Yet, as the majority
of these systems do not have an analytic solution, it is difficult
to quantify the accuracy of the algorithm. We thus apply it to perform
numerical integration using the scheme suggested by
Ref. \cite{li_numerical_2007} as a proof-of-principle.

Note, however, that our method is not meant to be an efficient
algorithm for performing numerical integration. As pointed out in
Ref. \cite{li_numerical_2007}, there is a one-to-one correspondence
between numerical integration and simulating an Ising model when put
under the Wang-Landau sampling framework. This applies to our
algorithm too and as long as we choose an integrand that is continuous
within the interval $[y_{\min}, y_{\max}]$, it is equivalent to the
situation of having a continuous energy domain for a physical
system. Moreover, numerical integration is indeed a more stringent
test case for our algorithm (and other histogram MC methods such as
Wang-Landau sampling in general), because the ``density of states''
$g(y)$ is usually more rugged than the density of states of a real
physical system.

If one can find an expression for the normalized $g(y)$, which
measures the portion of the domain within interval $[a, b]$
corresponding to a certain value of $y$, then the integral
can be found by summing the ``rows'' up (multiplied by the value of
$y$) instead of the columns in the following manner:
\begin{equation}
%\begin{aligned}
I = \int_a^b y(x) dx = \int_{y_{\min}}^{y_{\max}} g(y) y dy .
% \sum_{x_i=a}^{b}y(x_i)\Delta x
% \sum_{y_i=y_{\min}}^{y_{\max}}g(y_i)y_i .
%\end{aligned}
\end{equation}

Note that $g(y)$ needs to be normalized such that
\begin{equation}
\int_{y_{\min}}^{y_{\max}} g(y) dy = b-a.
\end{equation}

\begin{figure}[ht!]
\centering
\includegraphics[height=2in, width=0.85\columnwidth]{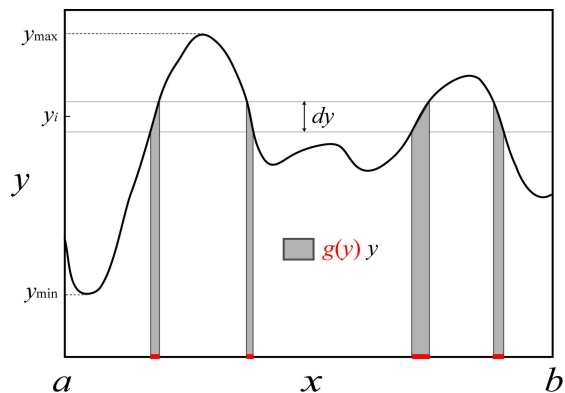}
\caption{A schematic diagram showing the numerical integration
  notations. The red regions on the $x$-axis marked the portion within
  the interval $[a, b]$ that gives a certain value $y$. All the areas
  shaded in grey add up to give $g(y)y$. }
\end{figure}

We apply our algorithm to perform the following integration where the
exact integral is known:
\begin{equation}
I = \int_{-2}^2 x^2 dx = \frac{16}{3} = 5.33333 \cdots ,
\end{equation}
and the ``density of states'' $g(y)$ can be expressed analytically:
\begin{equation}
g(y) = \frac{2(2-\sqrt{y})}{y}  \hspace{5mm}\textrm{for $y > 0$}.
\end{equation}

We use a Fourier sine series as the basis set $\{\psi(E)\}$ to fit the
remainder $R(E)$, and therefore a Fourier cosine series as the basis set
$\{\phi(E)\}$ for constructing the correction $\ln c(E)$ and updating
the density of states $\ln g(E)$. Moreover, we employ Kolmogorov-Smirnov
test in the $R(E)$ fitting step to see if the expression obtained is a
good fit to the dataset $\mathcal{D}$, using a criterion of $p =
0.5$. The experiment is done for different numbers of data in
the data set, with $k =$ 250, 500, 1000 and 2000.

\begin{figure}[h!]
\centering
\includegraphics[width=0.95\columnwidth]{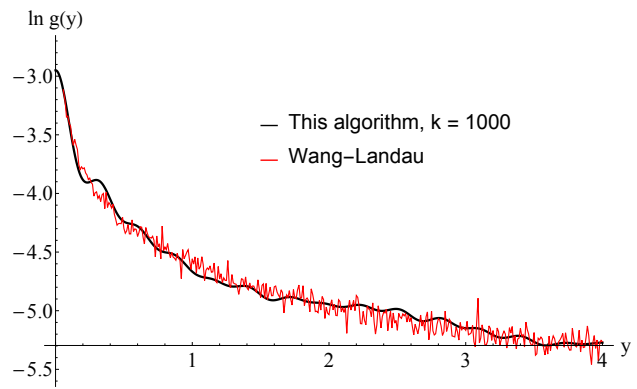}
\caption{Density of states $g(y)$ at the $120^\textrm{th}$ iteration,
  obtained using 1000 data points in a data set (black curve), a total
  of $1.2 \times 10^5$ MC steps are used. It is compared to a final
  $g(y)$ obtained using Wang-Landau sampling (red curve); this
  particular run requires $1.1 \times 10^6$ MC steps to complete. The
  DOS obtained from our algorithm is significantly smoother, yet its
  fluctuations fall within the statistical noise of the Wang-Landau
  DOS.}
\label{normDOS}
\end{figure}

We note that the Fourier sine and cosine series are not good
basis sets for this problem due to their oscillatory properties. Yet
the algorithm works surprisingly well. In Figure \ref{normDOS},  we
show a resulting density of states, $g(y)$, compared to that obtained
using Wang-Landau sampling. The fluctuations of our $g(y)$ fall within
the statistical noise of the WL density of states. 

\begin{figure}[ht!]
\centering
\includegraphics[width=0.95\columnwidth]{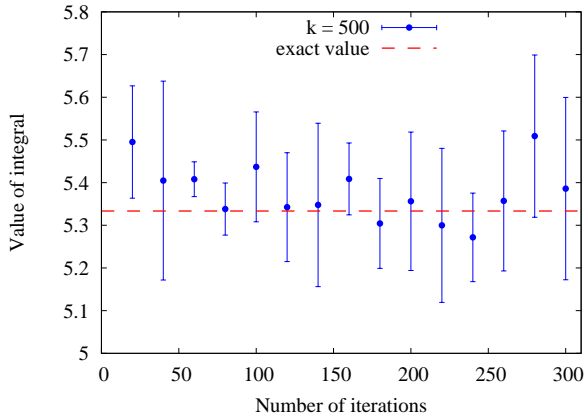}
\caption{Integral of $x^2$ over $x \in [-2,2]$, obtained using 500
  data points in a data set. Error bars are obtained from five
  independent runs.}
\label{k500}
\end{figure}

\begin{figure}[ht!]
\centering
\includegraphics[width=0.95\columnwidth]{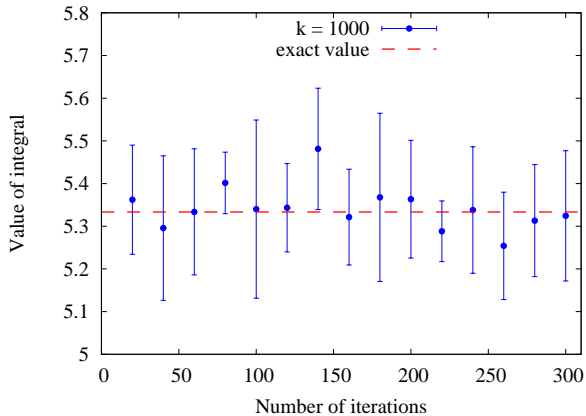}
\caption{Integral of $x^2$ over $x \in [-2,2]$, obtained using 1000
  data points in a data set. Error bars are obtained from five
  independent runs.}
\label{k1000}
\end{figure}

The values of the estimated integral at different iterations for $k =
500$ and $k = 1000$ are shown in Figures \ref{k500} and \ref{k1000},
respectively.

\begin{figure}[ht!]
\centering
\includegraphics[width=0.95\columnwidth]{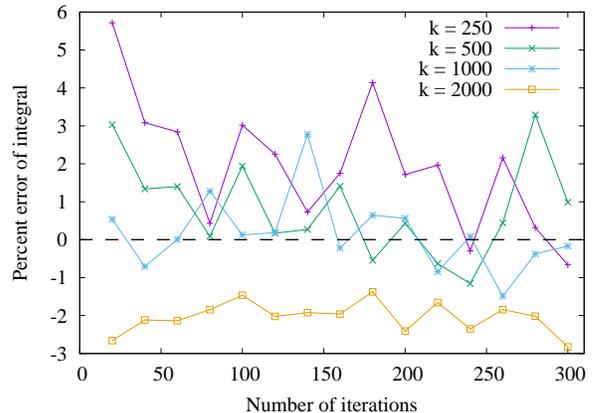}
\caption{Percent errors of the averaged estimated integrals at
  different iterations using various numbers of data points $k$ in a
  data set. The integrals for all cases are averaged over five
  independent runs. Except for $k=2000$, all other cases converge to
  the exact value eventually with small errors (within $\pm1\%$).}
\label{errors}
\end{figure}

We observe that the number of data points $k$ in a data set within an
iteration plays an important role in the accuracy. Both under-fitting
from insufficient data and over-fitting from excessive data would
produce inaccurate results. In both Figures \ref{k500} and
\ref{k1000}, most estimated integrals agree with the exact value to
within the error bars. No systematic correlation with the number of
iterations is observed for both the
estimated values of the integral and as well as the magnitude of the
error bars. Using $k=500$ or $k=1000$ does not seem to result in
significant  differences in the estimated value of the
integral. However, if we extend the studies and use fewer or more data
points in the data set $\mathcal{D}$, we observe different behavior as
shown in Figure \ref{errors}. For the $k = 250$ case, the integrals
are slightly overestimated at the first 200 iterations or so. The
percent errors fall back to within the same ranges as in the $k=500$ and
$k=1000$ cases later. This is reasonable because as the number of
iterations increases, more data are taken to correct the estimated
density of states.

However, the integrals are, unexpectedly, systematically
underestimated for the $k = 2000$ case. We also observe that the
number of terms in the expression of $R(E)$ and eventually $\ln g(E)$
generally increases with the number $k$ (Table \ref{nterms}).  A
larger number of data results in a more detailed fitting of the
DOS, hence more terms are used in the construction of the correction.
Unfortunately, there is also a higher risk of fitting the noise
``too well'', causing an over-fitting of the estimated DOS.  On the
other hand, using too few data points (such as $k = 250$) results in
larger fluctuations in the values of the intergral as well as in the
number of terms $N$ in the expression.  From our observations, using
about $k=1000$ data points in a data set is the safest and it strikes
a good balance between under-fitting (or even mal-fitting) and
overfitting.

\begin{table}
\centering
\begin{tabular}{|c|c|} \hline
$k$   &  $N$ \\ \hline
250   &  $19.4 \pm 14.6$ \\ \hline
500   &  $17.8 \pm 6.1$   \\ \hline
1000 &  $29.6 \pm 4.3$   \\ \hline
2000 &  $58.2 \pm 13.0$ \\
\hline\end{tabular}
\caption{\label{nterms} Averaged number of basis functions ($N$) for
  different number of data points ($k$) in each data set. The average
  values and errors are obtained from five independent runs.  The number
  $N$ is determined within the algorithm, at which the statistical
  test reaches $p \geq 0.5$. }
\end{table}

Note that the above experiments complete within hundreds of
iterations. Considering $k=1000$ data points in an iteration, the
total number of MC steps needed is of the order of
$10^5$. Comparing to the order of $10^6$ MC steps in Wang-Landau
sampling, our scheme is more efficient and it saves about 10$\times$
MC steps.  The reason is that when we correct the estimated DOS (i.e.,
sampling weights),  the correction is constructed to drive the random
walk \textit{intentionally} to achieve uniform sampling, or a ``flat
histogram'', as opposed to an incremental correction using the
histogram as in MUCA or WL sampling. We believe that our correction
scheme can also be applied to simple models with discrete energy
levels and yields significant speedup. 

\section{An improved scheme for better convergence}

While the results above showed that our proposed scheme is successful,
one problem is that it is still difficult to determine whether
convergence has been reached. Here, we suggest a possible way to
improve the quality of the results with two slight modifications to
the original scheme.

Firstly, when determining the number of terms for the
remainder $R(E)$ ($m$ in Eq. (\ref{R(E)})), the original scheme starts
from $m=1$ and increments it to $m+1$ sequentially until the
statistical test gives a score of $p \geq 0.5$. We observe that this
practice very often results in the update of the first few
coefficients only.  A remedy to it is that after the number $m$ is
determined in the first iteration, in the later iterations we propose
random permutations of the terms for the statistical test to start
with. This way, every coefficient will have a roughly equal chance to
get updated and refined.

Secondly, since the correction in Eq. (\ref{correction2}) will drive the
random walker in a way to achieve uniform sampling, we observe that it
is beneficial to use a milder correction update to drive the random
walker at a smaller step at a time. To do so, we rewrite Eq. (\ref{correction2})
with a pre-factor $s$ to take only a portion of $R(E)$ as
the correction:
\begin{equation}
\begin{aligned}
\label{correction3}
\ln c(E) &= s \sum_{i=1}^{m}r_i \frac{d\psi_i(E)}{dE}.
\end{aligned}
\end{equation}

\begin{figure}[ht!]
\centering
\includegraphics[width=0.95\columnwidth]{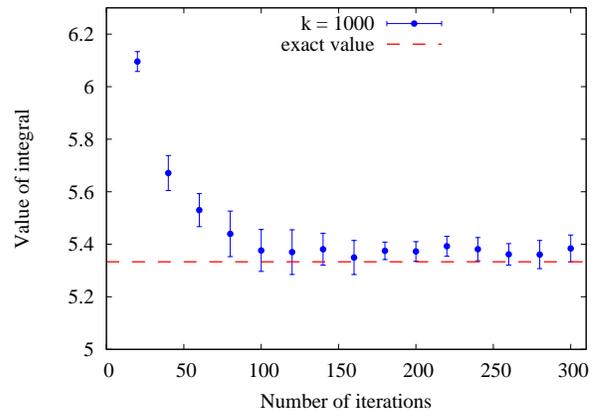}
\caption{Integral of $x^2$ over $x \in [-2,2]$, obtained using 1000
  data points in a data set and the improved scheme with a milder
  correction with the pre-factor $s = 0.25$. Error bars are obtained
  from five independent runs.}
\label{k1000_random}
\end{figure}

With these two small modifications, we revisited the integration problem
using $k = 1000$ data points in the data set (Figure
\ref{k1000_random}). The integral values in the first few dozens of
iterations deviate more from the exact value compared to the
original scheme, but it converges slowly to the exact value with a
much clearer convergence signal.  In this example, one may terminate
the simulation after e.g. the $150^{\textrm{th}}$ iteration. Another clear
improvement is that the error bar for each final answer is much reduced
compared to the original scheme, which indicates that the improved
scheme is able to give more precise results.

\section{Further improvements}

Our scheme is general in nature that the expression of the DOS $g(E)$
is not restricted to the Fourier form above, as long as there is a way
to formulate the correction $c(E)$ and the update formula. Obviously,
the quality of the resulting density of states depends heavily on the
choice of a proper basis set. While the plane wave basis used in the
present study allowed us to implement a prototype of our algorithm
without major effort, it suffers from serious issues that will require
the choice of more suitable basis sets. In particular, local
improvements to the density of states in a limited region of the
domain should not introduce changes in regions far away. Also, the
density of states often spans a wide range of values; indeed the
density of states for the integration example possesses a singularity
at the domain boundary. Thus a more suitable, localized, basis set
will greatly improve the convergence and accuracy of our method. 
 
Although we present our method as a serial algorithm so far, we stress
its parallelization is conceptually straight-forward. Since the
generation of data points (which is the energy evaluations for a
physical system) can be done independently by distributing the work
over different processors, simple ``poor-man's'' parallelization
strategy would already guarantee significant speedup in both strong
and weak scaling.

\section{Conclusions}

In this paper, we have presented a new Monte Carlo algorithm for
calculating probability densities of systems with a continuous energy domain. The idea is
inspired by combining ideas from the works of Wang and Landau
\cite{Wang2001a, Wang2001b}, Berg and Neuhaus \cite{berg_neuhaus_1991,
  berg_neuhaus_1992}, as well as Berg and Harris
\cite{berg_data_2008}. Nevertheless, our algorithm does not make use
of an explicit histogram as in traditional Wang-Landau or multicanonical
sampling. It is thus possible to avoid discrete binning of the
collected data. This histogram-free approach allows us to obtain the
estimated probability density, or the density of states, in terms of an
analytic expression.

We have demonstrated the application of our algorithm to a stringent
test case, numerical integration. Even with the sub-optimal Fourier
sine and cosine basis sets, our current algorithm is already capable
of giving reasonable results. An important point to note is that our
algorithm requires much \textit{fewer} number of Monte Carlo steps to
finish a simulation. It is enabled by the novel way we proposed to
correct the estimated density of states, and thus the sampling
weights, where the random walk is directed consciously to achieve
uniform sampling. This is essential for decreasing the
time-to-solution ratio of a simulation, especially for complex systems
where the computation time is dominated by energy evaluations.

The numerical integration test case provides useful insights into
improving the algorithm. Possible improvements include the use of a
basis set with local support and parallelization over energy
calculations. Our ongoing work includes all the possibilities for
perfecting the algorithm, as well as its application to simulations of
physical systems to solve real-world scientific problems.

\begin{acknowledgments}
The authors would like to thank T. W\"{u}st and D. M. Nicholson for
constructive discussions. This research was sponsored by the
Laboratory Directed Research and Development Program of Oak Ridge
National Laboratory, managed by UT-Battelle, LLC, for the
U. S. Department of Energy. This research used resources of the Oak
Ridge Leadership Computing Facility, which is supported by the Office
of Science of the U.S. Department of Energy under contract
no. DE-AC05-00OR22725.
\end{acknowledgments}

% Bibliography


\begin{thebibliography}{99}

\ifx \showCODEN    \undefined \def \showCODEN     #1{\unskip}     \fi
\ifx \showDOI      \undefined \def \showDOI       #1{#1}\fi
\ifx \showISBNx    \undefined \def \showISBNx     #1{\unskip}     \fi
\ifx \showISBNxiii \undefined \def \showISBNxiii  #1{\unskip}     \fi
\ifx \showISSN     \undefined \def \showISSN      #1{\unskip}     \fi
\ifx \showLCCN     \undefined \def \showLCCN      #1{\unskip}     \fi
\ifx \shownote     \undefined \def \shownote      #1{#1}          \fi
\ifx \showarticletitle \undefined \def \showarticletitle #1{#1}   \fi
\ifx \showURL      \undefined \def \showURL       {\relax}        \fi
% invisible to TeX
\providecommand\bibfield[2]{#2}
\providecommand\bibinfo[2]{#2}
\providecommand\natexlab[1]{#1}
\providecommand\showeprint[2][]{arXiv:#2}

\bibitem[\protect\citeauthoryear{Bennett}{Bennett}{1976}]%
        {BENNETT1976}
\bibfield{author}{\bibinfo{person}{Charles~H Bennett}.}
  \bibinfo{year}{1976}\natexlab{}.
\newblock \showarticletitle{Efficient estimation of free energy differences
  from {M}onte {C}arlo data}.
\newblock \bibinfo{journal}{{\it J. Comput. Phys.}} \bibinfo{volume}{22},
  \bibinfo{number}{2} (\bibinfo{year}{1976}), \bibinfo{pages}{245 -- 268}.
\newblock
\showISSN{0021-9991}
\showDOI{%
\url{https://doi.org/10.1016/0021-9991(76)90078-4}}


\bibitem[\protect\citeauthoryear{Berg and Harris}{Berg and Harris}{2008}]%
        {berg_data_2008}
\bibfield{author}{\bibinfo{person}{Bernd~A. Berg} {and}
  \bibinfo{person}{Robert~C. Harris}.} \bibinfo{year}{2008}\natexlab{}.
\newblock \showarticletitle{From data to probability densities without
  histograms}.
\newblock \bibinfo{journal}{{\em Computer Physics Communications\/}}
  \bibinfo{volume}{179}, \bibinfo{number}{6} (\bibinfo{date}{Sept.}
  \bibinfo{year}{2008}), \bibinfo{pages}{443--448}.
\newblock
\showISSN{0010-4655}
\showDOI{%
\url{https://doi.org/10.1016/j.cpc.2008.03.010}}


\bibitem[\protect\citeauthoryear{Berg and Neuhaus}{Berg and Neuhaus}{1991}]%
        {berg_neuhaus_1991}
\bibfield{author}{\bibinfo{person}{Bernd~A. Berg} {and} \bibinfo{person}{Thomas
  Neuhaus}.} \bibinfo{year}{1991}\natexlab{}.
\newblock \showarticletitle{Multicanonical algorithms for first order phase
  transitions}.
\newblock \bibinfo{journal}{{\em Physics Letters B\/}} \bibinfo{volume}{267},
  \bibinfo{number}{2} (\bibinfo{year}{1991}), \bibinfo{pages}{249 -- 253}.
\newblock
\showISSN{0370-2693}
\showDOI{%
\url{https://doi.org/10.1016/0370-2693(91)91256-U}}


\bibitem[\protect\citeauthoryear{Berg and Neuhaus}{Berg and Neuhaus}{1992}]%
        {berg_neuhaus_1992}
\bibfield{author}{\bibinfo{person}{Bernd~A. Berg} {and} \bibinfo{person}{Thomas
  Neuhaus}.} \bibinfo{year}{1992}\natexlab{}.
\newblock \showarticletitle{Multicanonical ensemble: A new approach to simulate
  first-order phase transitions}.
\newblock \bibinfo{journal}{{\em Phys. Rev. Lett.\/}}  \bibinfo{volume}{68}
  (\bibinfo{date}{Jan} \bibinfo{year}{1992}), \bibinfo{pages}{9--12}.
\newblock
Issue 1.
\showDOI{%
\url{https://doi.org/10.1103/PhysRevLett.68.9}}


\bibitem[\protect\citeauthoryear{Eisenbach, Zhou, Nicholson, Brown, Larkin, and
  Schulthess}{Eisenbach et~al\mbox{.}}{2009}]%
        {Eisenbach2009a}
\bibfield{author}{\bibinfo{person}{Markus Eisenbach},
  \bibinfo{person}{Cheng-Gang Zhou}, \bibinfo{person}{Don~M. Nicholson},
  \bibinfo{person}{Greg Brown}, \bibinfo{person}{Jeff Larkin}, {and}
  \bibinfo{person}{Thomas~C. Schulthess}.} \bibinfo{year}{2009}\natexlab{}.
\newblock \showarticletitle{A Scalable Method for Ab Initio Computation of Free
  Energies in Nanoscale Systems}. In \bibinfo{booktitle}{{\em Proceedings of
  the Conference on High Performance Computing Networking, Storage and
  Analysis}} {\em (\bibinfo{series}{SC '09})}. \bibinfo{publisher}{ACM},
  \bibinfo{address}{New York, NY, USA}, \bibinfo{pages}{64:1--64:8}.
\newblock


\bibitem[\protect\citeauthoryear{Ferrenberg and Swendsen}{Ferrenberg and
  Swendsen}{1989}]%
        {Ferrenber1989}
\bibfield{author}{\bibinfo{person}{Alan~M. Ferrenberg} {and}
  \bibinfo{person}{Robert~H. Swendsen}.} \bibinfo{year}{1989}\natexlab{}.
\newblock \showarticletitle{Optimized {M}onte {C}arlo data analysis}.
\newblock \bibinfo{journal}{{\em Phys. Rev. Lett.\/}}  \bibinfo{volume}{63}
  (\bibinfo{date}{Sep} \bibinfo{year}{1989}), \bibinfo{pages}{1195--1198}.
\newblock
Issue 12.
\showDOI{%
\url{https://doi.org/10.1103/PhysRevLett.63.1195}}


\bibitem[\protect\citeauthoryear{Hohenberg and Kohn}{Hohenberg and
  Kohn}{1964}]%
        {hohenberg_1964}
\bibfield{author}{\bibinfo{person}{Pierre Hohenberg} {and}
  \bibinfo{person}{Walter Kohn}.} \bibinfo{year}{1964}\natexlab{}.
\newblock \showarticletitle{Inhomogeneous Electron Gas}.
\newblock \bibinfo{journal}{{\em Phys. Rev.\/}} \bibinfo{volume}{136},
  \bibinfo{number}{3B} (\bibinfo{date}{Nov.} \bibinfo{year}{1964}),
  \bibinfo{pages}{B864--B871}.
\newblock
\showDOI{%
\url{https://doi.org/10.1103/PhysRev.136.B864}}


\bibitem[\protect\citeauthoryear{Hohenberg and Halperin}{Hohenberg and
  Halperin}{1977}]%
        {Hohenberg1977}
\bibfield{author}{\bibinfo{person}{Pierre~C. Hohenberg} {and}
  \bibinfo{person}{Bertrand~I. Halperin}.} \bibinfo{year}{1977}\natexlab{}.
\newblock \showarticletitle{Theory of dynamic critical phenomena}.
\newblock \bibinfo{journal}{{\em Rev. Mod. Phys.\/}}  \bibinfo{volume}{49}
  (\bibinfo{date}{Jul} \bibinfo{year}{1977}), \bibinfo{pages}{435--479}.
\newblock
Issue 3.
\showDOI{%
\url{https://doi.org/10.1103/RevModPhys.49.435}}


\bibitem[\protect\citeauthoryear{Hukushima and Nemoto}{Hukushima and
  Nemoto}{1996}]%
        {Hukushima1996}
\bibfield{author}{\bibinfo{person}{Koji Hukushima} {and} \bibinfo{person}{Koji
  Nemoto}.} \bibinfo{year}{1996}\natexlab{}.
\newblock \showarticletitle{Exchange {M}onte {C}arlo method and application to
  spin glass simulations}.
\newblock \bibinfo{journal}{{\em Journal of the Physical Society of Japan\/}}
  \bibinfo{volume}{65}, \bibinfo{number}{6} (\bibinfo{year}{1996}),
  \bibinfo{pages}{1604--1608}.
\newblock


\bibitem[\protect\citeauthoryear{Khan and Eisenbach}{Khan and
  Eisenbach}{2016}]%
        {Khan2016}
\bibfield{author}{\bibinfo{person}{Suffian~N. Khan} {and}
  \bibinfo{person}{Markus Eisenbach}.} \bibinfo{year}{2016}\natexlab{}.
\newblock \showarticletitle{Density-functional {M}onte-{C}arlo simulation of
  {C}u{Z}n order-disorder transition}.
\newblock \bibinfo{journal}{{\em Phys. Rev. B\/}}  \bibinfo{volume}{93}
  (\bibinfo{date}{Jan} \bibinfo{year}{2016}), \bibinfo{pages}{024203}.
\newblock
Issue 2.
\showDOI{%
\url{https://doi.org/10.1103/PhysRevB.93.024203}}


\bibitem[\protect\citeauthoryear{Kohn and Sham}{Kohn and Sham}{1965}]%
        {kohn_1965}
\bibfield{author}{\bibinfo{person}{Walter Kohn} {and} \bibinfo{person}{Lu~Jeu
  Sham}.} \bibinfo{year}{1965}\natexlab{}.
\newblock \showarticletitle{Self-Consistent Equations Including Exchange and
  Correlation Effects}.
\newblock \bibinfo{journal}{{\em Phys. Rev.\/}} \bibinfo{volume}{140},
  \bibinfo{number}{4A} (\bibinfo{date}{Nov.} \bibinfo{year}{1965}),
  \bibinfo{pages}{A1133--A1138}.
\newblock
\showDOI{%
\url{https://doi.org/10.1103/PhysRev.140.A1133}}


\bibitem[\protect\citeauthoryear{Landau and Binder}{Landau and Binder}{2015}]%
        {Landau_Binder}
\bibfield{author}{\bibinfo{person}{David~P. Landau} {and} \bibinfo{person}{Kurt
  Binder}.} \bibinfo{year}{2015}\natexlab{}.
\newblock \bibinfo{booktitle}{{\em A Guide to Monte Carlo Simulations in
  Statistical Physics, 4th Edition.}}
\newblock \bibinfo{publisher}{Cambridge University Press},
  \bibinfo{address}{Cambridge, U.K.}
\newblock


\bibitem[\protect\citeauthoryear{Li, W\"{u}st, Landau, and Lin}{Li
  et~al\mbox{.}}{2007}]%
        {li_numerical_2007}
\bibfield{author}{\bibinfo{person}{Ying~Wai Li}, \bibinfo{person}{Thomas
  W\"{u}st}, \bibinfo{person}{David~P. Landau}, {and} \bibinfo{person}{Hai-Qing
  Lin}.} \bibinfo{year}{2007}\natexlab{}.
\newblock \showarticletitle{Numerical integration using {W}ang{-}{L}andau
  sampling}.
\newblock \bibinfo{journal}{{\em Computer Physics Communications\/}}
  \bibinfo{volume}{177}, \bibinfo{number}{6} (\bibinfo{date}{Sept.}
  \bibinfo{year}{2007}), \bibinfo{pages}{524--529}.
\newblock
\showISSN{0010-4655}
\showDOI{%
\url{https://doi.org/10.1016/j.cpc.2007.06.001}}


\bibitem[\protect\citeauthoryear{Metropolis, Rosenbluth, Rosenbluth, Teller,
  and Teller}{Metropolis et~al\mbox{.}}{1953}]%
        {Metropolis1953}
\bibfield{author}{\bibinfo{person}{Nicholas Metropolis},
  \bibinfo{person}{Arianna~W. Rosenbluth}, \bibinfo{person}{Marshall~N.
  Rosenbluth}, \bibinfo{person}{Augusta~H. Teller}, {and}
  \bibinfo{person}{Edward Teller}.} \bibinfo{year}{1953}\natexlab{}.
\newblock \showarticletitle{Equation of State Calculations by Fast Computing
  Machines}.
\newblock \bibinfo{journal}{{\em The Journal of Chemical Physics\/}}
  \bibinfo{volume}{21}, \bibinfo{number}{6} (\bibinfo{year}{1953}),
  \bibinfo{pages}{1087--1092}.
\newblock
\showDOI{%
\url{https://doi.org/10.1063/1.1699114}}


\bibitem[\protect\citeauthoryear{Swendsen and Wang}{Swendsen and Wang}{1986}]%
        {Swendsen1986}
\bibfield{author}{\bibinfo{person}{Robert~H. Swendsen} {and}
  \bibinfo{person}{Jian-Sheng Wang}.} \bibinfo{year}{1986}\natexlab{}.
\newblock \showarticletitle{Replica {M}onte {C}arlo Simulation of
  Spin-Glasses}.
\newblock \bibinfo{journal}{{\em Phys. Rev. Lett.\/}}  \bibinfo{volume}{57}
  (\bibinfo{date}{Nov} \bibinfo{year}{1986}), \bibinfo{pages}{2607--2609}.
\newblock
Issue 21.
\showDOI{%
\url{https://doi.org/10.1103/PhysRevLett.57.2607}}


\bibitem[\protect\citeauthoryear{Torrie and Valleau}{Torrie and
  Valleau}{1977}]%
        {Torrie1977}
\bibfield{author}{\bibinfo{person}{G.M. Torrie} {and} \bibinfo{person}{J.P.
  Valleau}.} \bibinfo{year}{1977}\natexlab{}.
\newblock \showarticletitle{Nonphysical sampling distributions in {M}onte
  {C}arlo free-energy estimation: Umbrella sampling}.
\newblock \bibinfo{journal}{{\it J. Comput. Phys.}} \bibinfo{volume}{23},
  \bibinfo{number}{2} (\bibinfo{year}{1977}), \bibinfo{pages}{187 -- 199}.
\newblock
\showISSN{0021-9991}
\showDOI{%
\url{https://doi.org/10.1016/0021-9991(77)90121-8}}


\bibitem[\protect\citeauthoryear{Vogel, Li, W\"{u}st, and Landau}{Vogel
  et~al\mbox{.}}{2013}]%
        {Vogel2013}
\bibfield{author}{\bibinfo{person}{Thomas Vogel}, \bibinfo{person}{Ying~Wai
  Li}, \bibinfo{person}{Thomas W\"{u}st}, {and} \bibinfo{person}{David~P.
  Landau}.} \bibinfo{year}{2013}\natexlab{}.
\newblock \showarticletitle{Generic, Hierarchical Framework for Massively
  Parallel {Wang}-{L}andau Sampling}.
\newblock \bibinfo{journal}{{\em Phys. Rev. Lett.\/}}  \bibinfo{volume}{110}
  (\bibinfo{date}{May} \bibinfo{year}{2013}), \bibinfo{pages}{210603}.
\newblock


\bibitem[\protect\citeauthoryear{Vogel, Li, W\"{u}st, and Landau}{Vogel
  et~al\mbox{.}}{2014}]%
        {Vogel2014}
\bibfield{author}{\bibinfo{person}{Thomas Vogel}, \bibinfo{person}{Ying~Wai
  Li}, \bibinfo{person}{Thomas W\"{u}st}, {and} \bibinfo{person}{David~P.
  Landau}.} \bibinfo{year}{2014}\natexlab{}.
\newblock \showarticletitle{Scalable replica-exchange framework for
  {W}ang-{L}andau sampling}.
\newblock \bibinfo{journal}{{\em Phys. Rev. E\/}}  \bibinfo{volume}{90}
  (\bibinfo{date}{Aug} \bibinfo{year}{2014}), \bibinfo{pages}{023302}.
\newblock
Issue 2.
\showDOI{%
\url{https://doi.org/10.1103/PhysRevE.90.023302}}


\bibitem[\protect\citeauthoryear{Wang and Landau}{Wang and Landau}{2001a}]%
        {Wang2001b}
\bibfield{author}{\bibinfo{person}{Fugao Wang} {and} \bibinfo{person}{David~P.
  Landau}.} \bibinfo{year}{2001}\natexlab{a}.
\newblock \showarticletitle{Determining the density of states for classical
  statistical models: A random walk algorithm to produce a flat histogram}.
\newblock \bibinfo{journal}{{\em Phys. Rev. E\/}}  \bibinfo{volume}{64}
  (\bibinfo{date}{Oct} \bibinfo{year}{2001}), \bibinfo{pages}{056101}.
\newblock
Issue 5.
\showDOI{%
\url{https://doi.org/10.1103/PhysRevE.64.056101}}


\bibitem[\protect\citeauthoryear{Wang and Landau}{Wang and Landau}{2001b}]%
        {Wang2001a}
\bibfield{author}{\bibinfo{person}{Fugao Wang} {and} \bibinfo{person}{David.~P.
  Landau}.} \bibinfo{year}{2001}\natexlab{b}.
\newblock \showarticletitle{Efficient, Multiple-Range Random Walk Algorithm to
  Calculate the Density of States}.
\newblock \bibinfo{journal}{{\em Phys. Rev. Lett.\/}}  \bibinfo{volume}{86}
  (\bibinfo{date}{Mar} \bibinfo{year}{2001}), \bibinfo{pages}{2050--2053}.
\newblock
Issue 10.
\showDOI{%
\url{https://doi.org/10.1103/PhysRevLett.86.2050}}


\bibitem[\protect\citeauthoryear{Yin and Landau}{Yin and Landau}{2012}]%
        {Yin2012}
\bibfield{author}{\bibinfo{person}{Junqi Yin} {and} \bibinfo{person}{David~P.
  Landau}.} \bibinfo{year}{2012}\natexlab{}.
\newblock \showarticletitle{Massively parallel {W}ang-{L}andau sampling on
  multiple \{GPUs\}}.
\newblock \bibinfo{journal}{{\em Computer Physics Communications\/}}
  \bibinfo{volume}{183}, \bibinfo{number}{8} (\bibinfo{year}{2012}),
  \bibinfo{pages}{1568 -- 1573}.
\newblock
\showISSN{0010-4655}
\showDOI{%
\url{https://doi.org/10.1016/j.cpc.2012.02.023}}


\bibitem[\protect\citeauthoryear{Zierenberg, Marenz, and Janke}{Zierenberg
  et~al\mbox{.}}{2013}]%
        {Zierenberg2013}
\bibfield{author}{\bibinfo{person}{Johannes Zierenberg},
  \bibinfo{person}{Martin Marenz}, {and} \bibinfo{person}{Wolfhard Janke}.}
  \bibinfo{year}{2013}\natexlab{}.
\newblock \showarticletitle{Scaling properties of a parallel implementation of
  the multicanonical algorithm}.
\newblock \bibinfo{journal}{{\em Computer Physics Communications\/}}
  \bibinfo{volume}{184}, \bibinfo{number}{4} (\bibinfo{year}{2013}),
  \bibinfo{pages}{1155 -- 1160}.
\newblock
\showISSN{0010-4655}
\showDOI{%
\url{https://doi.org/10.1016/j.cpc.2012.12.006}}


\end{thebibliography}
\end{document}